\documentclass[a4paper,11pt]{article}

\usepackage{jinstpub} 
\usepackage{subfig}
\pdfoutput=1

\usepackage{lineno}
\usepackage{caption}
\title{A new indirect measurement method of the electron temperature for the Proto-sphera's pinch plasma}

\author[a]{D. Iannarelli,}
\author[b]{F. Napoli,}
\author [b]{F. Alladio,}
\author[b]{G. Apruzzese,}
\author[b]{F. Bombarda,}
\author [b]{P. Buratti,}
\author [c] {J. Delfini,}
\author [b]{A. De Ninno,}
\author[b]{F. Filippi,}
\author[b]{D. Fiorucci,}
\author[a]{A. Ingenito,}
\author [b]{S. Mannori,}
\author [b]{P. Micozzi,}
\author[a]{P. Teofilatto}

\affiliation[a]{Sapienza, School of Aerospace Engineering, Via Salaria 851-881, 00138 Rome, Italy}
\affiliation[b]{ENEA, Fusion and Nuclear Safety Department, Via E. Fermi 45, 00044 Frascati, Italy}
\affiliation[c]{Sapienza, Faculty of Mathematical, Physical and Natural Sciences, Piazzale Aldo Moro 5, 00185 Rome, Italy}

\emailAdd{daniele.iannarelli@uniroma1.it}
\emailAdd{francesco.napoli@enea.it}

\abstract{ This article presents a new method for estimating the electron temperature of the Proto-sphera’s screw pinch. The temperature radial profile is obtained by a self-consistent modeling of a 1D MHD equilibrium along with a 0D power balance of the plasma column, given measurements and estimates of the axial pinch plasma current, of the plasma rotational frequency and, at the equatorial plane, of the electron density radial profile, of the edge poloidal magnetic field, of the edge electron temperature and of the neutrals pressure in the vacuum vessel. The plasma is considered in equilibrium with its neutral phase and in constant rotation. A MATLAB code has been developed with the aim of estimating the MHD radial equilibrium profiles, the thermodynamic plasma state and the neutrals profile. The numerical estimates are compared with available experimental data showing a good agreement. }


\proceeding{6$^{\text{th}}$ International Conference Frontiers in Diagnostic Technology\\
  from 19/10/2022 to 21/10/2022\\
  Frascati (Rome), Italy}

\begin{document}
\maketitle
\flushbottom

\section{Introduction}
\label{sec:Introduction}

The Proto-sphera experiment is a plasma physics experiment of interest for the study of magnetically confined plasmas located in the Frascati Research Center (ENEA) which is capable of confining and sustaining a plasma in a spherical magnetic configuration. The plasma generated from the ionization of $H_{2}$, $He$, $Ar$, with only one gas injected each time, is ohmically heated by an electric discharge established between two electrodes. The overall magnetic configuration is the result of the magnetic fields generated by the plasma current, from external poloidal coils and from magnetic reconnections. A spherical magnetic configuration can be obtained without the metallic centerpost of a spherical tokamak making the design of the machine simpler and less demanding for maintenance. The plasma generated and confined inside Proto-sphera has temperatures of thousands of Kelvin and high electron densities of $10^{19}$-$10^{20}$ $m^{-3}$ and therefore is of potential interest for nuclear fusion research. This paper focuses on the estimation of the internal plasma state of the Proto-sphera’s pinch ~\cite{1}, ~\cite{2}, obtained modeling the thermodynamic state of the plasma column at the equatorial plane of the magnetic configuration where it can be probed by the installed diagnostics and where the plasma column can be better approximated as a screw pinch with constant cross-section.

\section{Physical background and mathematical model}

\begin{figure}[t!]
	\centering 
	\subfloat[]{\includegraphics[width =8.5 cm,height = 7.9 cm]{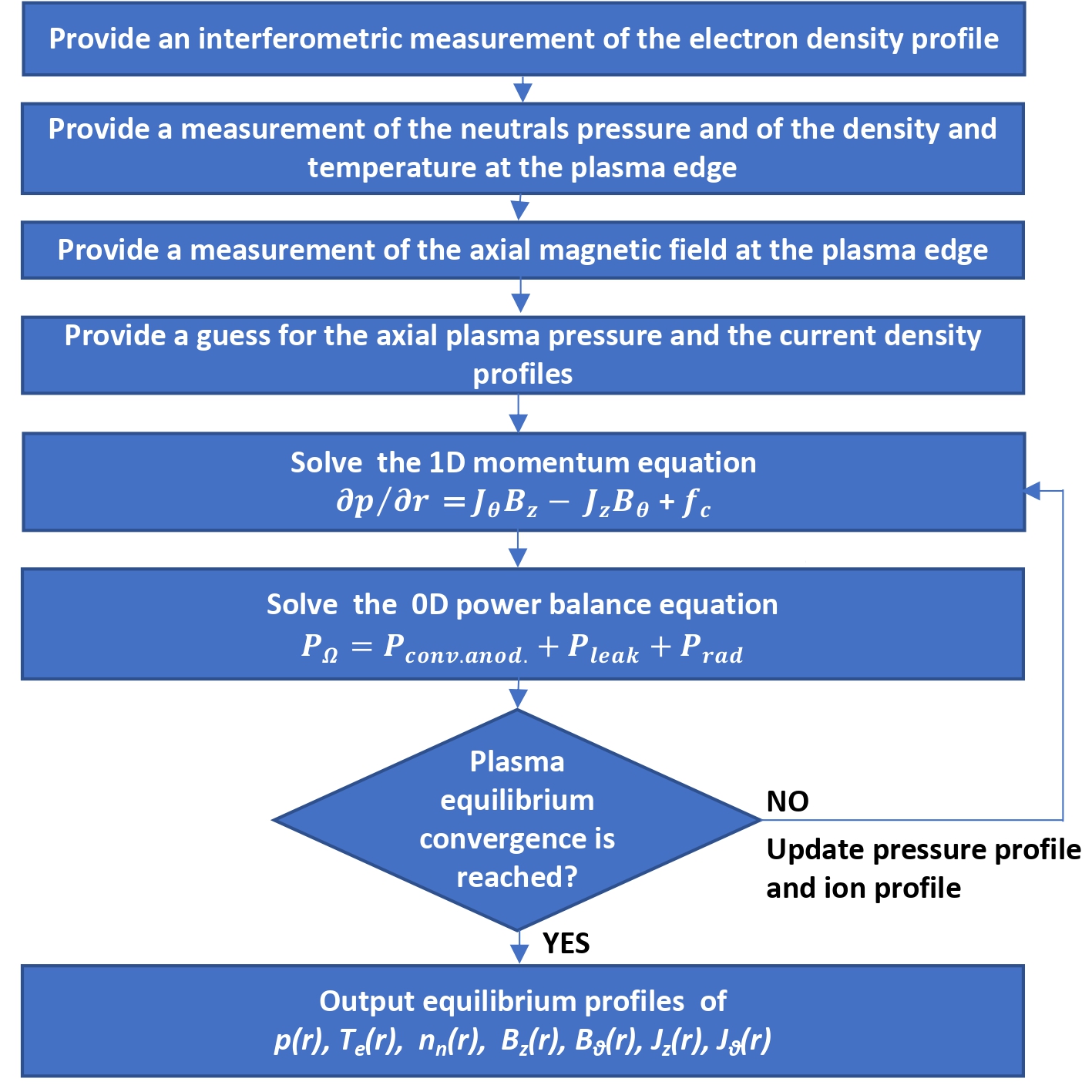}}
	\hspace{2 cm}
	\subfloat[]{\includegraphics[width = 4 cm, height = 7.55 cm]{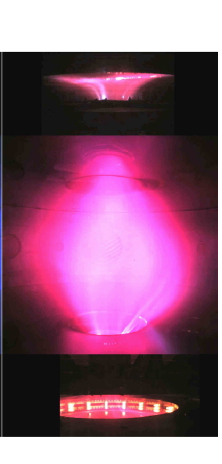}} 
	\caption{(a) Flowchart that describes the iterative procedure devised for the estimation of the electron pinch temperature. In each iteration the pressure and the ion profiles (derived by the implemented neutrals model) are updated according to the electron temperature profile derived by the solution of the power balance equation. (b) Proto-Sphera's pinch discharge in Hydrogen plasma.}
	\label{Figure 1}
\end{figure}

In this work the Proto-sphera’s pinch is modeled as a cylindrical plasma column with constant cross-section with an axial radially inhomogeneous plasma current flowing throught it and with an azimuthal magnetic field consistent with the equilibrium current profile. A poloidal external magnetic field is added to the azimuthal magnetic field to stabilize the central plasma column. In fact, the axial component of the magnetic field twists the field lines around the arc discharge enhancing the confinement of the central plasma column, which is therefore modeled as a screw pinch ~\cite{3}. The plasma is considered as a stationary conductive fluid in MHD equilibrium, in presence of a neutral gas of the same specie and with a constant angular rotation. Assuming an axial symmetry, an ideal radial (1D) MHD equilibrium model is coupled with a scalar (0D) power balance equation of the plasma pinch and the two models are solved self-consistently by an iterative procedure until numerical convergence is reached. Plasma pressure, electron density and axial current density radial profiles are assumed of exponential type, $p(r) = p_{0}\exp({-\alpha \hspace{1 mm} r^{2}/a^{2}}), \hspace{1 mm}  n_{e}(r) = n_{e0}\exp({-\beta \hspace{1 mm} r^{2}/a^{2}})$ and $J_{z} (r) = J_{z0}\exp({-\gamma \hspace{1 mm} r^{2}/a^{2}})$ where $p_{0}$, $n_{e0}$ and $J_{z0}$ are the profiles' peak values of the pressure, the electron density and the axial current density, $\alpha$, $\beta$ and $\gamma$ are the shaping values of the same profiles and $a$ is the plasma radius. Instead the temperature radial profile is not fixed but computed from the ratio of pressure and density profiles.
 Moreover, from the conservation of mass, neglecting the convective derivatives since we consider the plasma far from the walls of the vacuum vessel, we estimate the neutrals fraction inside the plasma as a balance between the ionization and the recombination of the species inside the plasma ($\Gamma_{n}^{ric} = \Gamma_{i}^{ion} $):

\[  n_{n} = n_{e}\frac{<\sigma_{ric}v_{e}>}{<\sigma_{ion} v_{e}>},  \quad for \hspace{1mm} signly \hspace{1mm} ionized \hspace{1mm}  particles   \]

\[  n_{n} = (\frac{n_{e} \frac{<\sigma_{ric}v_{e}>}{<\sigma_{ion}v_{e}>}_{1^{+}}  }{\frac{<\sigma_{ric}v_{e}>}{<\sigma_{ion}v_{e}>}_{2^{+}} + 2})(\frac{<\sigma_{ric}v_{e}>}{<\sigma_{ion}v_{e}>}_{2^{+}}),  \quad for\hspace{1mm}  double \hspace{1mm}  ionized \hspace{1mm}  particles  \]

where the ionization and the recombination rates are computed according to an electron impact ionization model as done in ~\cite{4}, ~\cite{5}.
According to the conservation of momentum, the variation of the momentum is equal to the net force applied to the plasma fluid. The external force acting on charged particles is the Lorentz force that is generated from the interaction between the charged particles and the magnetic field. Thus, the equation of conservation of momentum in the radial direction for the screw pinch reads as: 

\[  \frac{d}{dr}(\frac{B_{z}^{2}}{2\mu_{0}}) = - \frac{d}{dr}(p + \frac{B_{\theta}^{2}}{2\mu_{0}}) - \frac{B_{\theta}^{2}}{\mu_{0}r} + f_{c}  \]. 

where the above equation is solved for the Bz field component given the current density profile (fixed as a guess profile), the plasma pressure profile $p$ (derived iteratively by the power balance equation), the azimuthal field component $B_{\theta}$ (derived by the axial plasma current) and the centrifugal force $f_{c}$ (derived iteratively by the ion profile). 
According to the conservation of energy, the variation in time of the internal energy of the plasma pinch must be the difference between the heating power and the power lost from the plasma. At steady state an energetic equilibrium is reached when the heating ohmic power ($P_{\Omega}$) equals the total power losses, modeled as the sum of the electron convective losses to the anode ($P_{conv.anode} $~\cite{6}), the radiative power losses emitted by the plasma ($P_{rad}$ ~\cite{7}, ~\cite{8})  and the power lost in the ejection of the charged particles in the loss cone of the magnetic configuration ($P_{leak}$ ~\cite{6}):

\[  P_{\Omega}  = P_{conv. anode} + P_{leak} + P_{rad}         \]

where the above equation is solved in the mean electronic temperature to update iteratively the pressure profile of the MHD equilibrium computation.
A flowchart of the mathematical and physical model that is coded in MATLAB to obtain the pinch's equilibrium is reported in Figure ~\ref{Figure 1} (a).

\section{Results and conclusions}

\begin{figure}[htbp]
\centering 
\includegraphics[width=0.4\textwidth]{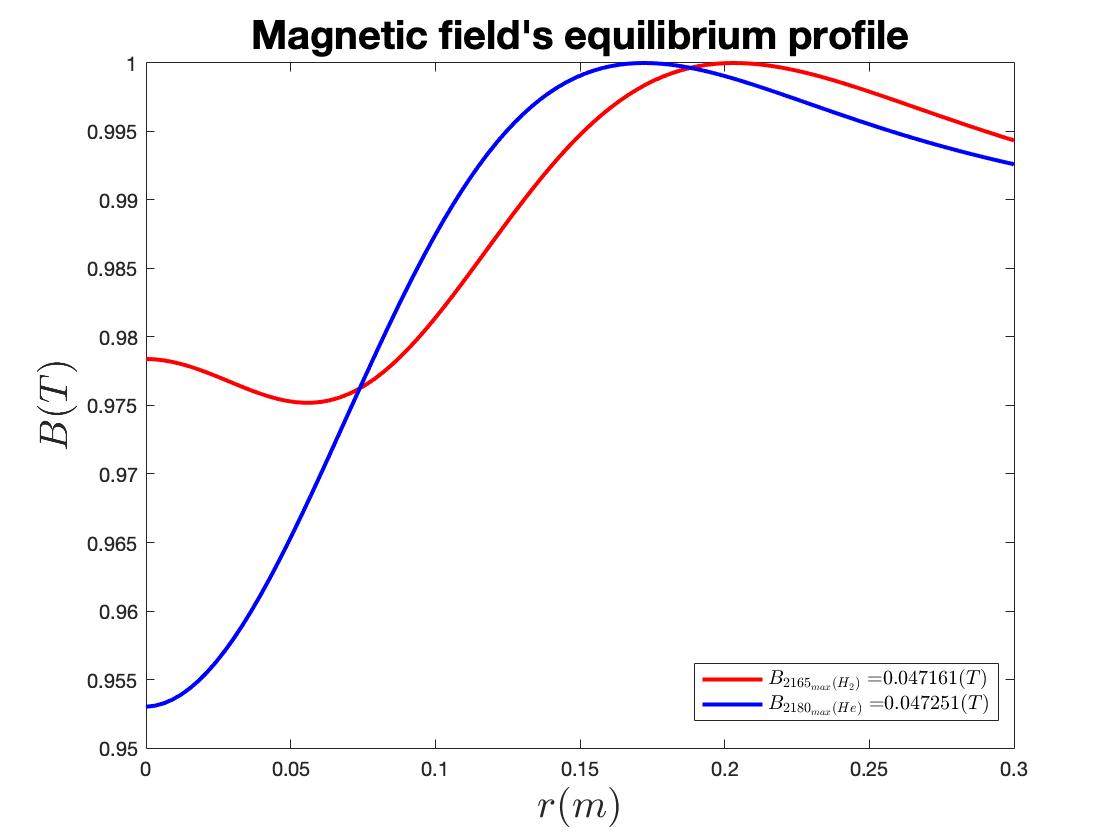}
\includegraphics[width=0.4\textwidth]{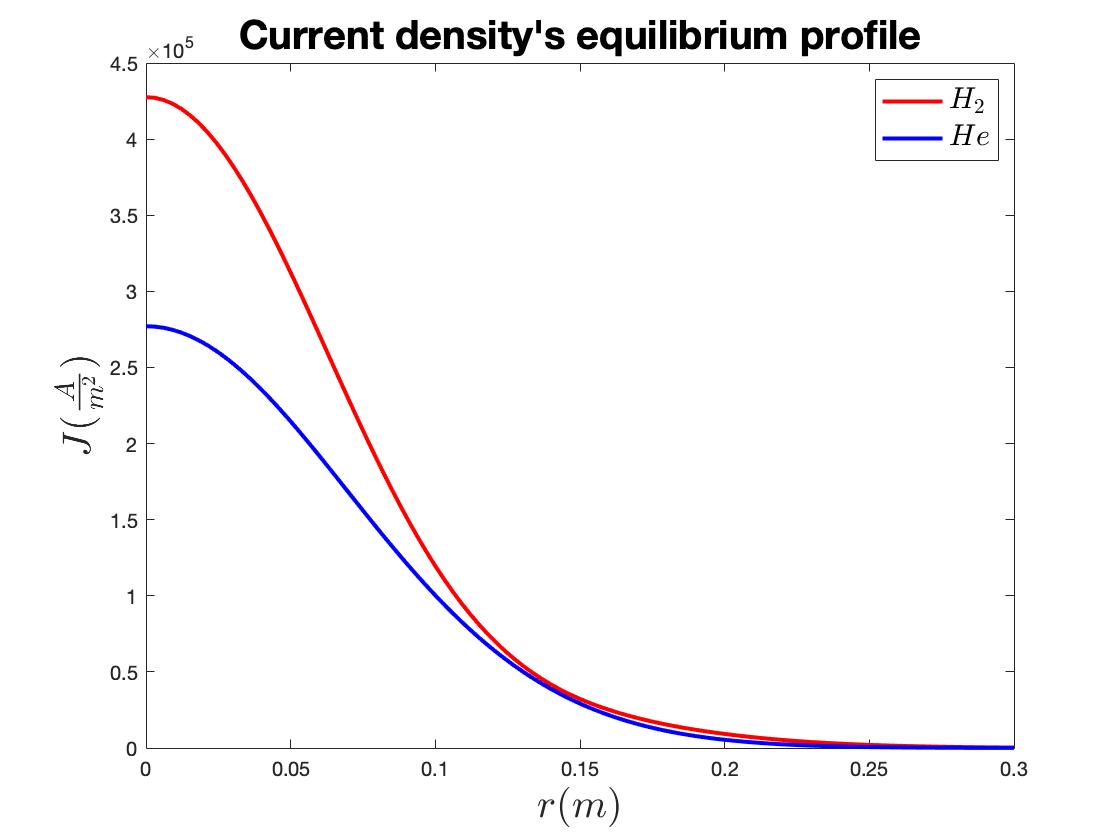}

\includegraphics[width=0.4\textwidth]{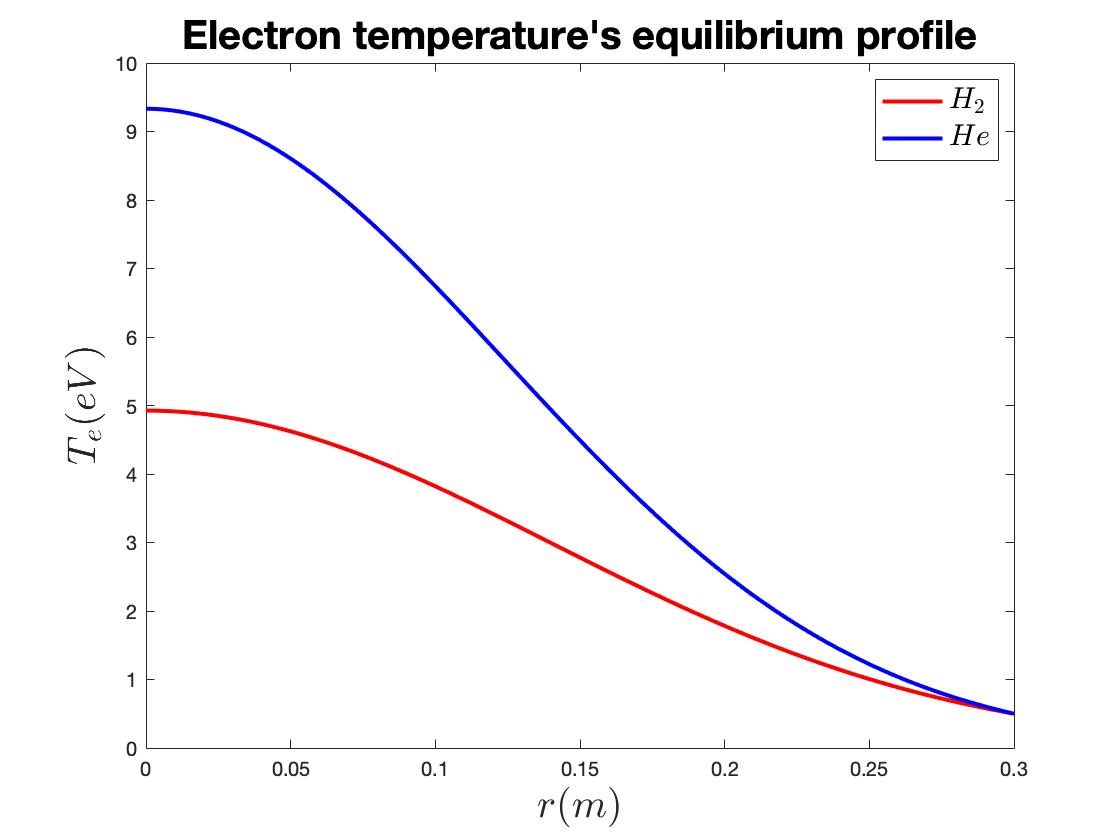}
\includegraphics[width=0.4\textwidth]{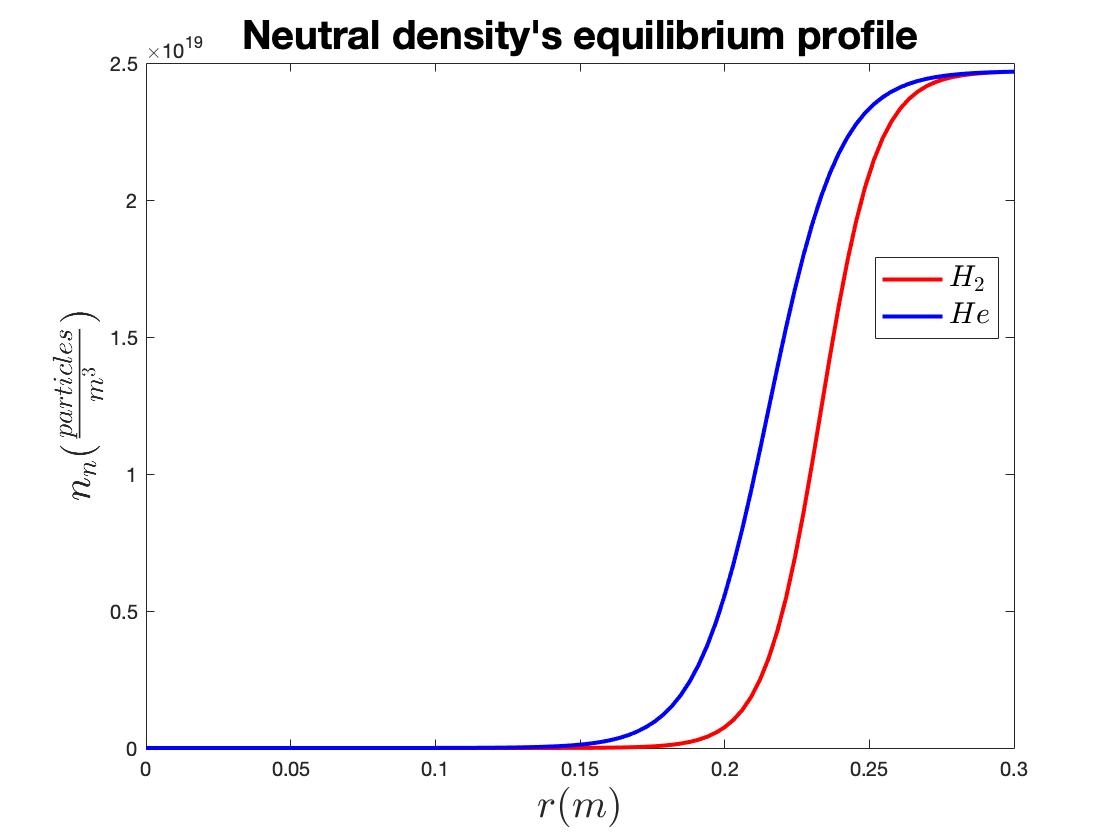}

\caption{ Plasma pinch's equilibrium profiles}
\label{Equilibrium profiles}
\end{figure}

A MATLAB code has been written which is capable to simulate the radial equilibrium profiles of the Proto-sphera’s pinch (Figure ~\ref{Figure 1} (b)). We derive the input electron density profile of exponential type from measurements of the electron density at the plasma edge measured by a Langmuir probe and of the line average electron density measured by interferometry. In Figure ~\ref{Equilibrium profiles} are reported the equilibrium profiles obtained for Hydrogen and Helium, where we considered respectively a pinch axial current of 10.50 kA and 8.50 kA, a line average electronic density of 4.18 10\textsuperscript{19} m\textsuperscript{-3} and 1.80 10\textsuperscript{19} m\textsuperscript{-3}, an edge poloidal magnetic field of  4.69 10\textsuperscript{-2} T and 5.86 10\textsuperscript{-2} T, an edge electron density of 2 10\textsuperscript{18} m\textsuperscript{-3} and 3 10\textsuperscript{17} m\textsuperscript{-3}, and a small concentration of impurities (Nitrogen and Oxigen) and, for both cases, an edge electron temperature of 0.50 eV, a plasma radius of 0.30 m, a rotational frequency of 100 Hz,  and a neutrals pressure of 0.10 Pa considered at a constant room temperature of 293.15 K.

To evaluate our estimates, a comparison is made with the experimental data as reported in table  ~\ref{Table 1}. In Proto-sphera, temperature measurements are obtained for Helium only with a spectroscopic technique ~\cite{9} and radiation power measurements are made with a diamond detector sensitive to the UV radiation  ~\cite{10}.  Our estimations for the mean electronic temperature and radiated power are in a good agreement with measurements.

\begin{table}[htpb]
	\centering
	\begin{tabular}{|c|c|c|c|c|c|c|}
		\hline
		Specie&Shot&$I_{z} (kA)$&$T_{e_{num}} (eV) $&$T_{e_{exp}} (eV)$& $P_{rad_{num}} (kW) $& $P_{rad_{exp}} (kW) $  \\
		\hline
		$H_{2}$&2165&10.5&3.32& N.A.&42.0&42.7  \\
		\hline 
		He&2180&8.5&6.21& 5.5* &46.2&43.9  \\
		\hline 
	\end{tabular}
	\caption{Comparison between the numerical data (indicated with the $num$ subscript) and the experimental data (indicated with the $exp$ subscript). The value denoted by asterisk * refers to a reference value taken from the shot 1910.}
	\label{Table 1}
\end{table}

\end{document}